\documentclass[]{spie}  %>>> use for US letter paper
\usepackage{amsmath,amsfonts,amssymb}
\usepackage{graphicx}
\usepackage{multirow}
\usepackage[colorlinks=true, allcolors=blue]{hyperref}

\title{Patch Stitching Data Augmentation for Cancer Classification in Pathology Images}

\author[]{Jiamu Wang}
\author[]{Chang-Su Kim}
\author[*]{Jin Tae Kwak}
\affil[]{School of Electrical Engineering, Korea University, Seoul 02841, Korea}
\authorinfo{$^*$ corresponding author: e-mail: jkwak@korea.ac.kr, telephone: +82 2 3290-3241; fax 82 2 921-0544; kwaklab.net}

\begin{document} 
\maketitle

\begin{abstract}
Computational pathology, integrating computational methods and digital imaging, has shown to be effective in advancing disease diagnosis and prognosis. In recent years, the development of machine learning and deep learning has greatly bolstered the power of computational pathology. However, there still remains the issue of data scarcity and data imbalance, which can have an adversarial effect on any computational method. In this paper, we introduce an efficient and effective data augmentation strategy to generate new pathology images from the existing pathology images and thus enrich datasets without additional data collection or annotation costs. To evaluate the proposed method, we employed two sets of colorectal cancer datasets and obtained improved classification results, suggesting that the proposed simple approach holds the potential for alleviating the data scarcity and imbalance in computational pathology.
\end{abstract}

% Include a list of keywords after the abstract 

\section{Introduction}
\label{sec:intro}  % \label{} allows reference to this section
Computational pathology is an emerging practice that involves computational processing and analysis of digitized tissue specimens with respect to the presence, progression, and outcome of various diseases \cite{cui2021artificial}. Owing to the recent development of computational methods, in particular machine learning and deep learning, and the easy access and availability of digital scanners, numerous computational pathology tools have been proposed and shown to be effective in expediting the workflow of pathology and improving diagnostic and prognostic accuracy \cite{huang2021integration} \cite{li2023self}. However, computational pathology, in general, still suffers from a lack of sufficient data or annotations. Although there is a growing number of pathology datasets, data imbalance is another hurdle that limits the full potential of deep learning models in computational pathology and thus hampers the translation of such models to clinics. 

To address these challenges, data augmentation has emerged as a pivotal technique. Traditional augmentation methods such as image flipping, rotation, resizing, and cropping have been widely adopted in computational pathology. Several advanced approaches have been recently proposed in machine learning and computer vision communities and have demonstrated their efficacy in enhancing classification tasks; for example, CutOut \cite{devries2017improved} randomly selects and masks out regions in an image, MixUp \cite{zhang2017mixup} combines two random images in the alpha channel, and CutMix \cite{yun2019cutmix} randomly replaces a region in one image with another region in the other image.

Herein, we propose a simple yet effective data augmentation method, named as the $\bold{Pa}$tch $\bold{S}$titching image $\bold{S}$ynthesis (PaSS), to improve the accuracy and robustness of deep learning models in computational pathology. PaSS creates new images by combining segments from various existing pathology images within the same pathological category via two distinct clipping techniques, resulting in PaSS$_{Rec}$ and PaSS$_{SLIC}$. 
%We showcase the performance of these two methods in Fig.~\ref{fig:m1} and Fig.~\ref{fig:m2} and visually compare their differences in Fig.~\ref{fig:m3} and performance in Table~\ref{tab:tab1}. 
Two sets of colorectal cancer datasets are employed to test the efficacy and efficiency of the two methods. Without incurring additional costs for data acquisition or annotation, PaSS is able to improve the classification performance, demonstrating the potential in mitigating data shortages and imbalances in computational pathology.

\section{METHODS}

\subsection{Patch Stitching Image Synthesis}
\label{sec:title}

\begin{figure} [ht]
\begin{center}
\begin{tabular}{c} %% tabular useful for creating an array of images 
\includegraphics[height=6cm]{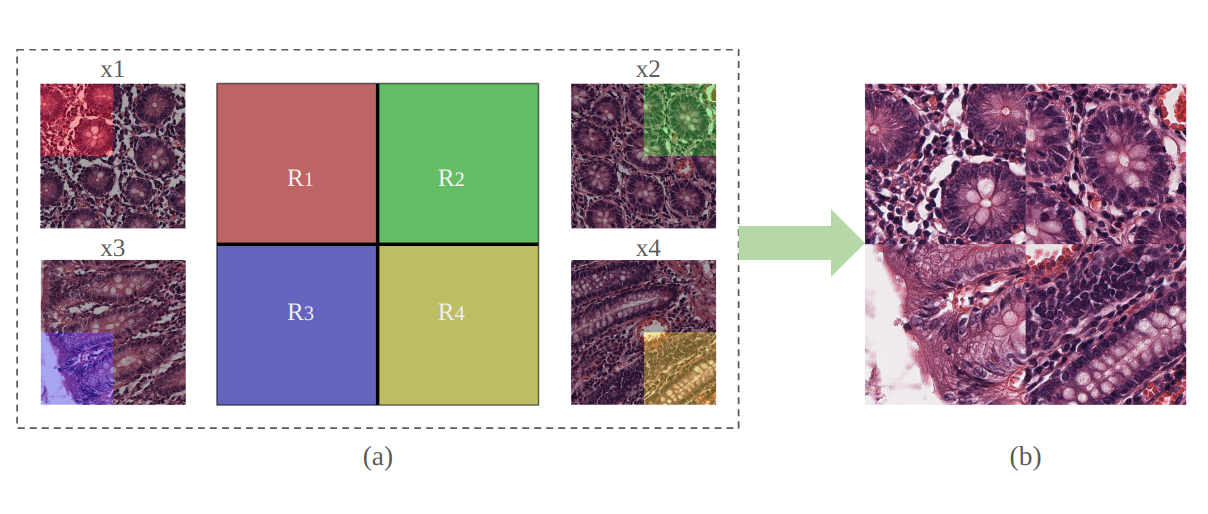}
\end{tabular}
\end{center}
\caption[example] 
%>>>> use \label inside caption to get Fig. number with \ref{}
{ \label{fig:m1} Illustration of PaSS$_{Rec}$ under $P$ = 4. (a) It divides an empty canvas into four square sections, crops the areas corresponding to each of these sections from four different images, and (b) combines these cropped sections from each image with its respective regions on the canvas to form the complete stitched image.}
\end{figure}

\begin{figure} [ht]
\begin{center}
\begin{tabular}{c} %% tabular useful for creating an array of images 
\includegraphics[height=9cm]{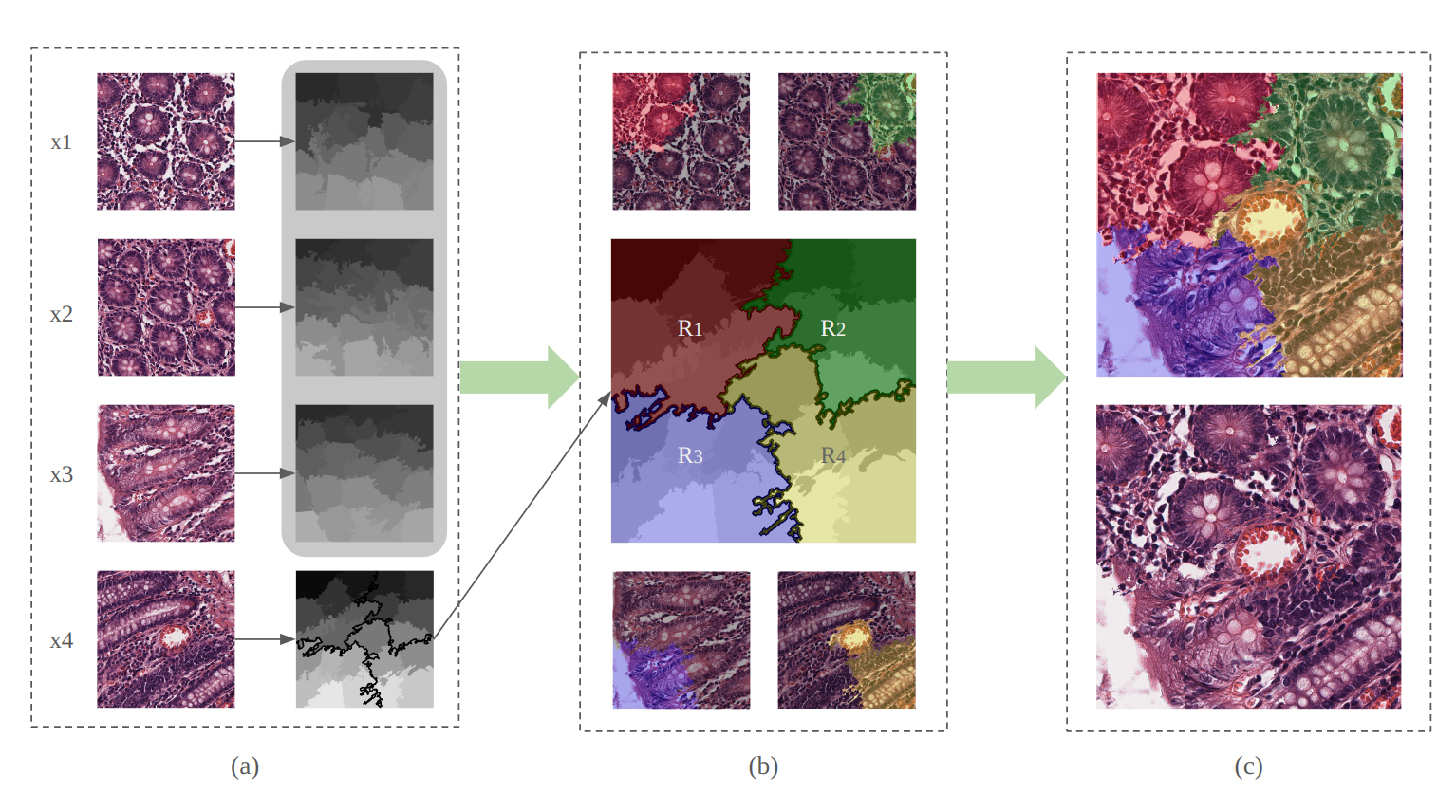}
\end{tabular}
\end{center}
\caption[example] 
%>>>> use \label inside caption to get Fig. number with \ref{}
{ \label{fig:m2} Illustration of PaSS$_{SLIC}$ under $P$ = 4. (a) It generates a superpixel map for each image and selects the image with the highest number of superpixels. (b) It combines these superpixels into four roughly equal regions and clips the corresponding sections from four different images. (c) It creates the stitched image by merging these clipped regions from each image with its respective area on the canvas.}
\end{figure}

Suppose that we are given a number of pathology images $\{x_i^c|i=1,..., N_c,c=1,..., C\}$ where $x_i^c$ denotes the $i$th pathology image in a pathology category $c$, $N_c$ is the number of pathology images in the category $c$, and $C$ represents the number of distinct pathology categories.
PaSS generates a set of new images per pathology category. For each category $c$, PaSS synthesizes a new pathology image $z$ in five steps: 1) It prepares a blank canvas $z$, 2) randomly picks $P$ pathology images $\{x_i^c|i=1,..., P\}$, 3) divides each pathology images into $P$ disjoint regions $R_{x_i} = \cup_{j=1}^{P} R_{{x_i},j}, i=1,..., C$ where $R_{{x_i},j}$ is the $j$th disjoint region in the pathology image $x_i^c$, 4) selects one image $x_k^c, 1 \le k \le P$ and use its disjoint regions $R_{x_k}$ to divide the blank canvas $z$, resulting in $R_{z}$, 5) takes each region $R_{z, i}$ from the corresponding image $x_{i}^c$, and pastes it into the blank canvas $z(R_{{z_i}, i}) = x_i^c(R_{{z_i}, i}), i=1,..., P$. 
In order to split each image into disjoint regions, we adopt the following two approaches: 1) PaSS$_{Rec}$: it divides each image into a number of rectangular regions (Fig.~\ref{fig:m1}) and randomly chooses one image to divide $z$; 2) PaSS$_{SLIC}$: it utilizes the SLIC algorithm \cite{achanta2012slic} to obtain a superpixel map, including the number of superpixels $\gg P$, for each image, selects the one image with the highest number of superpixels for dividing $z$, combines superpixels into $P$ disjoint regions whose size is roughly the same as shown in Fig.~\ref{fig:m2}.

% Note: If compiling with LaTeX+dvipdf, please ensure images generated from 
% other software packages have their bounding boxes set correctly.

% Note: If compiling with LaTeX+dvipdf, please ensure images generated from 
% other software packages have their bounding boxes set correctly.

\subsection{Dataset}
We employ two colorectal cancer datasets that are publicly available \cite{Vuong21}. The first dataset includes 9,857 image patches of size 1024 $\times$ 1024 pixels ($\sim$ 258 $\mu$m $\times$ 258 $\mu$m), digitized using an Aperio digital slide scanner (Leica Biosystems). These image patches are grouped into four categories, including 1,600 benign (BN), 2,322 well-differentiated (WD) cancer, 4,105 moderately-differentiated (MD) cancer, and 1,830 poorly-differentiated (PD) cancer, and are divided into a training set of 7,027 images ($Train$), a validation set of 1,242 images ($Validation$), and a test set of 1,588 images ($Test I$). The second dataset contains 110,170 image patches (27,986 BN, 8,394 WD, 61,985 MD, and 11,985 PD) of size 1144 $\times$ 1144 pixels ($\sim$ 258 $\mu$m $\times$ 258 $\mu$m), scanned using a NanoZoomer digital slide scanner (Hamamatsu Photonics K.K.). We note that the second dataset, designated as $Test II$, was acquired using a different digital slide scanner at different time periods from the first dataset. 

\subsection{Cancer Classification}
Using the colorectal cancer datasets, we conduct colorectal cancer classification using two deep convolutional neural networks (CNNs): 1) EfficientNet-B0 \cite{Tan19} and 2) ResNet-50 \cite{He16}. For each CNN model, we use $Train$ to optimize the weights of the model, select the best model using $Validation$, and test the best model on $Test I$ and $Test II$. 

In order to assess the effectiveness of PaSS, we generate $K$ synthesized images per pathology category using PaSS$_{Rec}$ and PaSS$_{SLIC}$ separately. We set $K$ to 600. The number of disjoint regions $P$ is set to 2, 4, and 9. In total, we generate 2,400 synthetic images per method. Then, the synthesized images are added to $Train$ and repeat the above experiment, i.e., only the number of images in $Train$ is increased and $Validation$, $Test I$, and $Test II$ are kept the same.

\subsection{Training Procedure}
All the image patches are resized to 512 $\times$ 512 pixels. We train all the models under consideration for 50 epochs with Adam optimizer (${\beta}_1 = 0.9, {\beta}_2 = 0.999, {\epsilon} = 1.0e^{-8}$) and cosine annealing warm restarts schedule \cite{Lo16} with an initial learning rate of $1.0e^{-4}$, $eta\_min=1.0e^{-5}$, and $T_0 = 25$, decreasing the learning rate to $1.0e^{-5}$ after 25 epochs and restarting the learning rate from $1.0e^{-4}$. 
All the models and methods are implemented on PyTorch. Both EfficientNet-B0 and ResNet-50 are initialized with ImageNet pre-trained weights.
Data augmentation techniques are listed as follows: 1) a random horizontal flip with a 50 \% chance; 2) a random vertical flip with a 50 \% chance; 3) one of Gaussian blurring, average blurring, or median blurring; 4) additive Gaussian noise independently per channel with a 50 \% chance 5) drop from 1 to 10 \%  pixels independently per channel with a 50 \% chance 6) color change in hue, saturation in the interval [-20, 20]; 7) modification of the contrast of images according to $127 + \alpha*(v-127)$, where $v$ is a pixel value and $\alpha$ is sampled uniformly from the interval [0.5, 2.0], independently per channel with a 50 \% chance. 1)and 2) are applied to every image patch. 3), 4), 5), 6), and 7) are executed from 0 to 5 to an image patch. All the augmentation techniques are implemented using Aleju library (https://github.com/aleju/imgaug).

\begin{table}[!ht]
\caption{Classification Scores under $P$ = 2, 4, 9} 
\label{tab:tab1}
\begin{center}       
\begin{tabular}{ccccccc} 
\hline
\rule[-1ex]{0pt}{3.5ex} \multirow{2}{*}{Model} & \multirow{2}{*}{Image Synthesis}  & \multirow{2}{*}{$P$} & \multicolumn{2}{c}{$Test I$}  & \multicolumn{2}{c}{$Test II$} \\ \cline{4-7}
\rule[-1ex]{0pt}{3.5ex}  & &  & Acc (\%) & F1 & Acc (\%) & F1 \\
\hline
\rule[-1ex]{0pt}{3.5ex} \multirow{7}{*}{EfficientNet-B0} & No & - & 87.47 & 0.8754 & 75.28 & 0.7764   \\
\cline{2-7}
\rule[-1ex]{0pt}{3.5ex}  & \multirow{3}{*}{PaSS$_{Rec}$} & 2 & 88.60 & 0.8841 & 75.97 & 0.7815    \\
\rule[-1ex]{0pt}{3.5ex}  &  & 4 & $\mathbf{88.79}$ & 0.8862 & 77.39 & 0.7958   \\
\rule[-1ex]{0pt}{3.5ex}  &  & 9 & $\mathbf{88.79}$ & $\mathbf{0.8870}$ & 76.59 & 0.7880    \\
\cline{2-7}
\rule[-1ex]{0pt}{3.5ex}  &  \multirow{3}{*}{PaSS$_{SLIC}$} & 2 & 88.29 & 0.8807 & 76.50 & 0.7860  \\
\rule[-1ex]{0pt}{3.5ex}  &  & 4 & 88.41 & 0.8812 & $\mathbf{78.53}$ & $\mathbf{0.8043}$  \\
\rule[-1ex]{0pt}{3.5ex}  &  & 9 & 88.16 & 0.8798 & 76.66 & 0.7892   \\
\hline
\rule[-1ex]{0pt}{3.5ex} \multirow{7}{*}{ResNet-50} & No & - & 85.52 & 0.8566 & 77.53 & 0.7952   \\
\cline{2-7}
\rule[-1ex]{0pt}{3.5ex}  & \multirow{3}{*}{PaSS$_{Rec}$} & 2 & $\mathbf{85.96}$ & 0.8561 & 78.33 & 0.7995    \\
\rule[-1ex]{0pt}{3.5ex}  &  & 4 & 85.39 & 0.8545 & 76.79 & 0.7879   \\
\rule[-1ex]{0pt}{3.5ex}  &  & 9 & 85.89 & 0.8570 & 79.42 & 0.8095    \\
\cline{2-7}
\rule[-1ex]{0pt}{3.5ex}  &  \multirow{3}{*}{PaSS$_{SLIC}$} & 2 & $\mathbf{85.96}$ & $\mathbf{0.8579}$ & 76.83 & 0.7852  \\
\rule[-1ex]{0pt}{3.5ex}  &  & 4 & 85.20 & 0.8503 & $\mathbf{80.11}$ & $\mathbf{0.8139}$  \\
\rule[-1ex]{0pt}{3.5ex}  &  & 9 & 85.52 & 0.8542 & 76.45 & 0.7830   \\
\hline
\end{tabular}
\end{center}
\end{table}

\section{Results}
Table \ref{tab:tab1} demonstrates the results of colorectal cancer classification using two CNN models including EfficientNet-B0 and ResNet-50. The classification performance was measured using two evaluation metrics: 1) accuracy (Acc): the fraction of correctly classified samples to the total number of samples and 2) weighted F1 score (F1): a weighted average of per-class harmonic means of precision and recall.
Without using the synthesized images, i.e., using the original images only, EfficientNet-B0 and ResNet-50 achieved 87.47\% Acc and 0.8754 F1 and 85.52\% and 0.8566 F1 on $Test I$, respectively. As tested on $Test II$, the performance of both EfficientNet-B0 and ResNet-50 was substantially dropped; for instance, 12.19\% Acc and $0.099$ F1 for EfficientNet-B0 and 7.99\% Acc and $0.0614$ F1 for ResNet-50. This may be ascribable to the difference in the image acquisition conditions or settings such as digital slide scanners. 

Using the synthesized images during training, there was, in general, a performance gain for both datasets. For example, the classification performance of EfficientNet-B0 was improved by $\le$1.32\% Acc and $\le$0.0116 F1 on $Test I$ and $\le$3.25\% Acc and $\le$0.0279 F1 on $Test II$. Similar observations were made for ResNet-50, achieving a performance gain of $\le$0.44\% Acc and $\le$0.013 F1 on $Test I$ and $\le$2.58\% Acc and $\le$0.0187 F1 on $Test II$. In a head-to-head comparison between the two stitching methodologies (PaSS$_{Rec}$ and PaSS$_{SLIC}$), differences between models and datasets were observed. On $Test I$ and $Test II$, the highest Acc and F1 were obtained by EfficientNet-B0 using Pass$_{Rec}$ with $P$=9 and ResNet-50 using PaSS$_{SLIC}$ with $P$=4, respectively. This is, by and large, attributable to the backbone network, i.e., EfficietNet-B0 and ResNet-50, since EfficientNet-B0 performs better on $Test I$ and ResNet-50 shows better performance on $Test II$. Moreover, for both models, the best performance on $Test II$ was obtained by using PaSS$_{SLIC}$ with $P$=4, suggesting the superior generalization ability of the method.

\begin{figure}[ht]
\begin{center}
\begin{tabular}{c} %% tabular useful for creating an array of images 
\includegraphics[height=11cm]{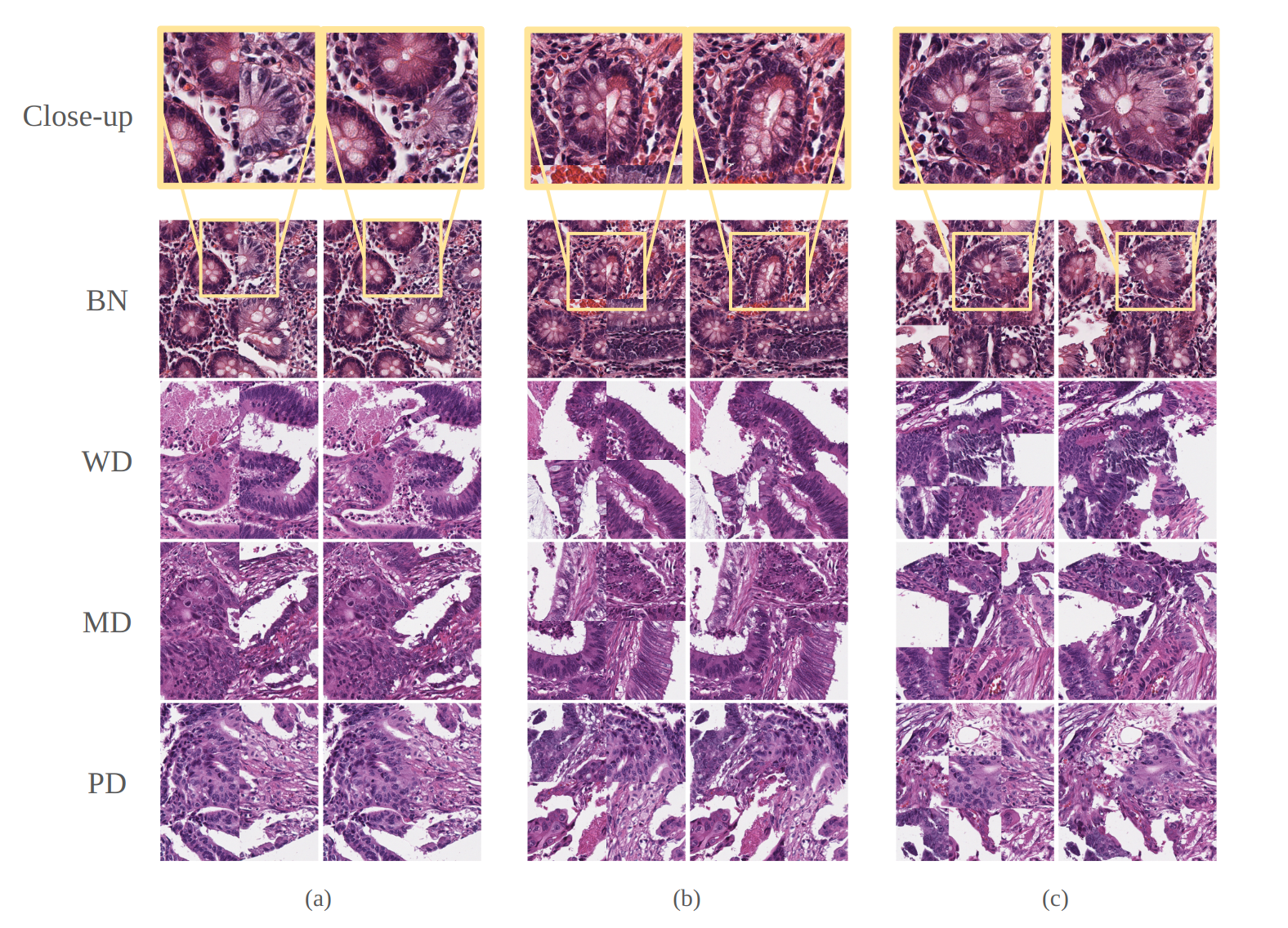}
\end{tabular}
\end{center}
\caption[example] 
%>>>> use \label inside caption to get Fig. number with \ref{}
{ \label{fig:m3}Visual comparison of PaSS$_{Rec}$(left) and PaSS$_{SLIC}$(right) under (a) $P$=2, (b) $P$=4, and (c) $P$=9.}
\end{figure}

Both PaSS$_{Rec}$ and PaSS$_{SLIC}$ mix patches in the spatial domain and incorporate a similar amount of pixels from other images in the same category. PaSS$_{SLIC}$, however, better preserves tissue structures by clipping them along the outlines given by superpixels, offering more realistic visual results, while PaSS$_{Rec}$ clips in the shape of rectangular, resulting in obvious clipping artifacts, as demonstrated in Fig.~\ref{fig:m3}.

% Using the synthesized images during training, there was, in general, a performance gain for both datasets. For example, the classification performance of EfficientNet-B0 was improved by $\le$1.19\% Acc and $\le$0.0098 F1 on $Test I$ and $\le$2.35\% Acc and $\le$0.0205 F1 on $Test II$. Similar observations were made for ResNet-50, achieving a performance gain of $\le$1.32\% Acc and $\le$0.0109 F1 on $Test I$ and $\le$1.45\% Acc and $\le$0.0114 F1 on $Test II$. In a head-to-head comparison between the two stitching methodologies (PaSS$_{Rec}$ and PaSS$_{Slic}$), differences between models and datasets were observed. On $Test I$ and $Test II$, the highest Acc and F1 were obtained by EfficientNet-B0 using Pass$_{Rec}$ with $P$=9 and ResNet-50 using PaSS$_{Rec}$ with $P$=4, respectively. This is, by and large, attributable to the backbone network, i.e., EfficietNet-B0 and ResNet-50, since EfficientNet-B0 performs better on $Test I$ and ResNet-50 shows better performance on $Test II$. On the other hand, ResNet-50 using PaSS$_{Rec}$ with $P$=2 and EfficietNet-B0 using PaSS$_{Slic}$ with $P$=4 achieved the largest performance gain on $Test I$ and $Test II$, respectively.

\section{CONCLUSIONS}
We introduce an approach of image synthesis to enhance the accuracy and robustness of pathology image classification. The approach generates a set of new pathology images from the existing images with known labels without posing high computational resources and costs. The experimental results confirm the feasibility of the proposed simple yet effective image synthesis method. The approach is generic and can be applied to any type of disease and tissue. In the follow-up study, we will further investigate the proposed approach to multi-organ datasets.

\section*{ACKNOWLEDGMENTS}
This study was supported by the National Research Foundation of Korea (NRF) grant funded by the Korean government (MSIP) (No. 2021R1A2C2014557 and No. 2021R1A4A1031864). 

% References
\bibliography{report} % bibliography data in report.bib

\begin{thebibliography}{10}

\bibitem{cui2021artificial}
Cui, M. and Zhang, D.~Y., ``Artificial intelligence and computational
  pathology,'' {\em Laboratory Investigation}~{\bf 101}(4),  412--422 (2021).

\bibitem{huang2021integration}
Huang, Z., Chai, H., Wang, R., Wang, H., Yang, Y., and Wu, H., ``Integration of
  patch features through self-supervised learning and transformer for survival
  analysis on whole slide images,'' in [{\em Medical Image Computing and
  Computer Assisted Intervention--MICCAI 2021: 24th International Conference,
  Strasbourg, France, September 27--October 1, 2021, Proceedings, Part VIII
  24}{\nolinebreak\hspace{0.1em}]},   561--570, Springer (2021).

\bibitem{li2023self}
Li, L., Liang, Y., Shao, M., Lu, S., Ouyang, D., et~al., ``Self-supervised
  learning-based multi-scale feature fusion network for survival analysis from
  whole slide images,'' {\em Computers in Biology and Medicine}~{\bf 153},
  106482 (2023).

\bibitem{devries2017improved}
DeVries, T. and Taylor, G.~W., ``Improved regularization of convolutional
  neural networks with cutout,'' {\em arXiv preprint arXiv:1708.04552}  (2017).

\bibitem{zhang2017mixup}
Zhang, H., Cisse, M., Dauphin, Y.~N., and Lopez-Paz, D., ``mixup: Beyond
  empirical risk minimization,'' {\em arXiv preprint arXiv:1710.09412}  (2017).

\bibitem{yun2019cutmix}
Yun, S., Han, D., Oh, S.~J., Chun, S., Choe, J., and Yoo, Y., ``Cutmix:
  Regularization strategy to train strong classifiers with localizable
  features,'' in [{\em Proceedings of the IEEE/CVF international conference on
  computer vision}{\nolinebreak\hspace{0.1em}]},   6023--6032 (2019).

\bibitem{achanta2012slic}
Achanta, R., Shaji, A., Smith, K., Lucchi, A., Fua, P., and S{\"u}sstrunk, S.,
  ``Slic superpixels compared to state-of-the-art superpixel methods,'' {\em
  IEEE transactions on pattern analysis and machine intelligence}~{\bf 34}(11),
   2274--2282 (2012).

\bibitem{Vuong21}
Vuong, T. T.~L., Kim, K., Song, B., and Kwak, J.~T., ``Joint categorical and
  ordinal learning for cancer grading in pathology images,'' {\em Medical Image
  Analysis} ,  73 (2021).

\bibitem{Tan19}
Tan, M. and Le, Q.~V., ``Efficientnet: rethinking model scaling for
  convolutional neural networks,'' {\em Proceedings of the 36th International
  Conference on Machine Learning} ,  6105--6114 (2019).

\bibitem{He16}
He, K., Zhang, X., Ren, S., and Sun, J., ``Deep residual learning for image
  recognition,'' {\em CVPR} ,  770--778 (2016).

\bibitem{Lo16}
Loshchilov, I. and Hutter, F., ``Sgdr: Stochastic gradient descent with warm
  restarts,'' {\em arXiv preprint arXiv:1608.03983}  (2016).

\end{thebibliography}
\bibliographystyle{spiebib} % makes bibtex use spiebib.bst

\end{document}